\documentclass[english, amsmath,amssymb, reprint, twocolumn, longbibliography]{revtex4-1}
\usepackage[T1]{fontenc}
\usepackage[latin9]{inputenc}
\usepackage{amsthm}
\usepackage{graphicx}
\usepackage{enumitem}
\usepackage{braket}
\usepackage{bbm}
\usepackage{bm}
\usepackage{mathtools}
\usepackage[unicode=true, breaklinks=false, pdfborder={0 0 1}, backref=false, colorlinks=true, linkcolor=blue, urlcolor=blue, citecolor=blue]{hyperref}

\makeatletter

\theoremstyle{plain}
\newtheorem{thm}{\protect\theoremname}
\theoremstyle{plain}
\newtheorem{lem}{\protect\lemmaname}
\theoremstyle{plain}
\newtheorem{cor}{\protect\corollaryname}
\theoremstyle{plain}
\newtheorem{prop}{\protect\propositionname}

\allowdisplaybreaks

\newcommand{\tr}{\operatorname{tr}}
\newcommand{\id}{\mathbbm{1}}

\let\originalleft\left
\let\originalright\right
\renewcommand{\left}{\mathopen{}\mathclose\bgroup\originalleft}
\renewcommand{\right}{\aftergroup\egroup\originalright}

\renewcommand\bra[1]{{\langle{#1}|}}
\makeatletter
\renewcommand\ket[1]{
  \@ifnextchar\bra{\k@t{#1}\!}{\k@t{#1}}
}
\newcommand\k@t[1]{{|{#1}\rangle}}
\makeatother

\usepackage{babel}
\providecommand{\corollaryname}{Corollary}
\providecommand{\lemmaname}{Lemma}
\providecommand{\propositionname}{Proposition}
\providecommand{\theoremname}{Theorem}

\begin{document}

\title{Quantum discord-breaking and discord-annihilating channels}

\author{Thao P. Le}
\email{thao.le.16@ucl.ac.uk}

\affiliation{Department of Physics and Astronomy, University College London, Gower
Street, London WC1E 6BT, United Kingdom}

\begin{abstract}
Quantum discord-breaking channels were previously defined as the local channels that act on subsystem $A$ to produce classical-quantum states across system $AB$. However, unlike entanglement, discord is asymmetric. Here, we characterise the discord-breaking channels that act on subsystem $B$, thus completing the overall description of discord-breaking channels. We then introduce the notion of discord-annihilating channels, which act globally on system $AB$ to destroy quantum discord, and find their closed form. Discord-annihilating channels have clear operational description, involving subspace projections on subsystem $A$ and conditional preparation of fixed states on subsystem $B$. 
\end{abstract}
\maketitle

\section{Introduction}

Quantum entanglement has been, and continues to be, the focus of much of quantum information theory \citep{Horodecki2009}. Entanglement is required in multiple quantum applications such as quantum key distribution and quantum computation \citep{Steane1998,Jennewein2000}. This has lead to the study of the mechanisms that hinder and destroy entanglement. For example, in dynamical processes, entanglement might undergo sudden death \citep{Yu2009} and/or a sudden birth \citep{Lopez2008}. In discrete settings, there have been investigations into the robustness of entanglement against added noise \citep{Vidal1999}, closed-form characterisations of entanglement-breaking quantum channels \citep{Ruskai2003,Horodecki2003} and explorations of entanglement-annihilating channels \citep{Moravcikova2010,Filippov2012,Filippov2013}. However, there are a number of nonclassical correlations beyond entanglement that also lead to non-trivial advantages in various quantum tasks: one of which is quantum discord.

Quantum discord measures the purely quantum correlations between systems \citep{Ollivier2001,Henderson2001,Modi2014,Modi2012a}. Discord can exist even in separable states, and is useful in tasks such as quantum metrology and parameter estimation \citep{Girolami2014,Micadei2015}. As such, the preservation of quantum discord is also integral to successful quantum applications. There are many analogous studies for quantum discord, from its robustness against noise and sudden death \citep{Werlang2009}, to the characterisation of discord-breaking channels \citep{Guo2013,Yao2013}. However, we argue that the prior characterisation of discord-breaking channels is incomplete.

Unlike separable states, zero-discord states are asymmetric and have a preferred subsystem: a system $AB$ can be classical-quantum (and hence zero-discord by one definition), yet \emph{not} quantum-classical. Prior work on discord-breaking channels \citep{Guo2013,Yao2013} only considered the set of local channels acting on the preferred subsystem $A$. In this paper, we consider the class of discord-breaking channels that act locally on the non-preferred system $B$. We find that these are exactly the fixed-point channels, thus completing the description of discord-breaking channels.

But then what of \emph{nonlocal} discord-breaking channels that act on the entire system $AB$? We call these \emph{discord-annihilating channels}---in analogy to entanglement-annihilating channels \citep{Moravcikova2010,Filippov2012,Filippov2013}. These channels destroy discord (or entanglement) \emph{within} the system they act upon, as opposed to destroying correlations \emph{between} the affected system and any external system. We find that discord-annihilating channels involve a combination of projective measurements on subsystem $A$ combined with conditional state preparation on subsystem $B$. We provide the explicit characterisation of these channels.

\begin{figure}
\begin{centering}
\includegraphics[width=1\columnwidth]{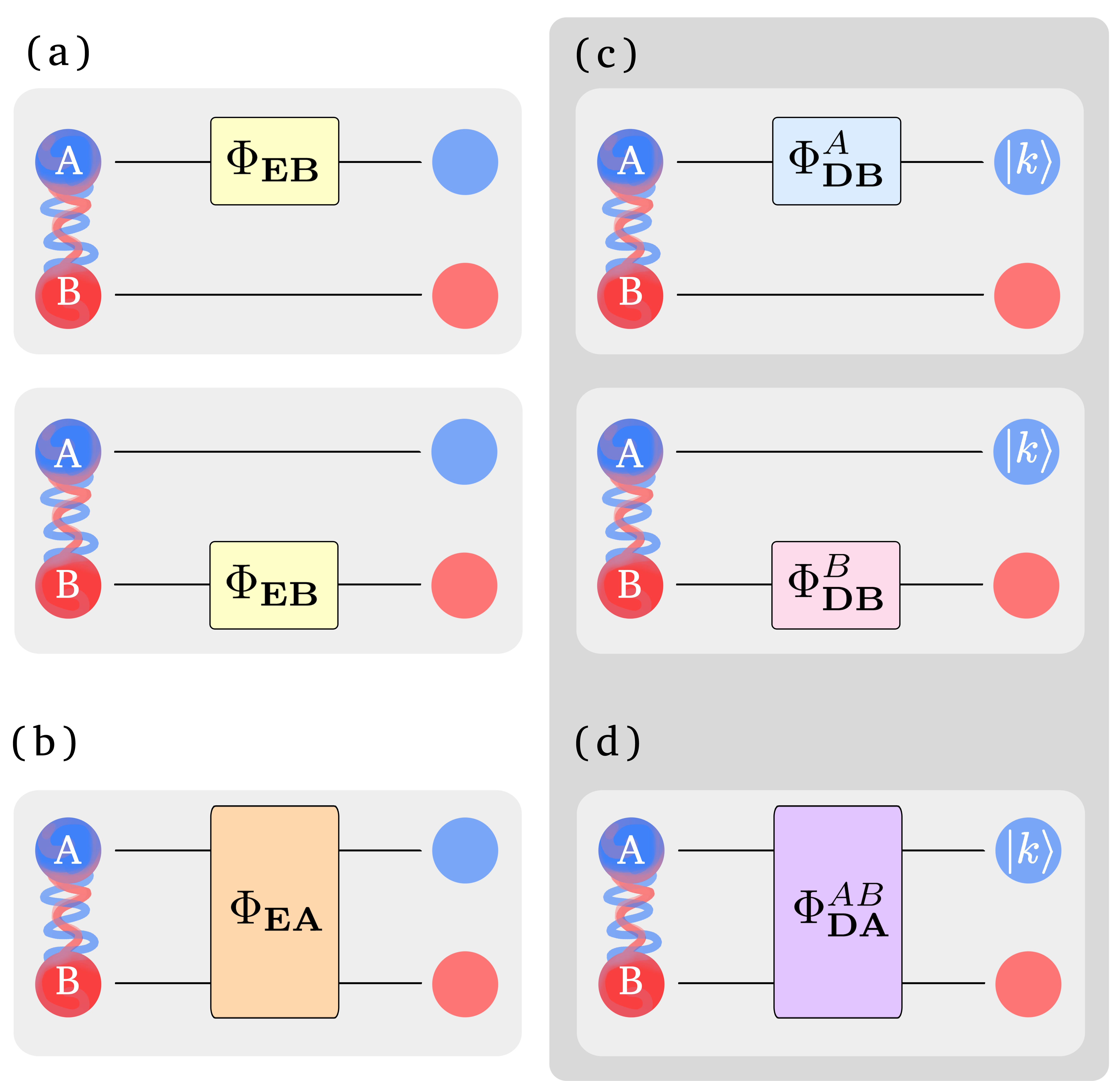}
\par\end{centering}
\caption{Various correlation destroying channels.
(a) Entanglement-breaking (EB) channels $\Phi_{\text{\textbf{EB}}}$ act locally and break entanglement with any external system \citep{Ruskai2003,Horodecki2003}. 
(b) Entanglement-annihilating (EA) channels $\Phi_{\text{\textbf{EA}}}$ act on a multipartite system and break entanglement among its subsystems \citep{Moravcikova2010,Filippov2012,Filippov2013}.
(c) Discord-breaking (DB) channels act upon one of the subsystems to break discord and produce classical-quantum states such that subsystem $A$ is diagonal in some basis $\{|k\rangle\}$. Type A channels, $\Phi_{\text{\textbf{DB}}}^{A}$ act on the first system, $A$ \citep{Yao2013,Guo2013}. Type B channels $\Phi_{\text{\textbf{DB}}}^{B}$ act on the second system $B$ (see Theorem \ref{thm:discord-breaking channels}).
(d) Discord-annihilating (DA) $\Phi_{\text{\textbf{DA}}}^{AB}$ channels act upon the bipartite system to break discord within the system (see Theorem \ref{thm:DA_channels}).\label{fig:EB_EA_DB_DA}}
\end{figure}

In Fig. \ref{fig:EB_EA_DB_DA}, we illustrate the actions of entanglement-breaking (EB), entanglement-annihilating (EA), discord-breaking (DB) and discord-annihilating (DA) channels. Entanglement-breaking channels already have a closed-form representation \citep{Ruskai2003,Horodecki2003}. As of the conclusion of this paper, discord-breaking and -annihilating channels will also closed-form representations. There is currently no general closed-form for entanglement-annihilating channels.

This paper is organised as follows. In Sec. \ref{sec:Preliminaries} we introduce some notation and preliminaries about entanglement, discord, and the breakings thereof. In Sec. \ref{sec:Discord-breaking-channels}, we describe the discord-breaking channels that act on subsystem $B$. In Sec. \ref{sec:Discord-annihilating-channels}, we investigate discord-annihilating channels. We conclude in Sec. \ref{sec:Conclusion}.

\section{Notation and Preliminaries\label{sec:Preliminaries}}

A quantum system $X$ has an associated Hilbert space $\mathcal{H}^{X}$. The set of states---density operators that are positive semidefinite and have unit trace---is denoted $\mathcal{S}\left(\mathcal{H}^{X}\right)$. For a bipartite system $AB$, the total Hilbert space is $\mathcal{H}^{AB}=\mathcal{H}^{A}\otimes\mathcal{H}^{B}$. Quantum channels are described as completely positive trace-preserving (CPTP) maps, $\Phi:\mathcal{L}\left(\mathcal{H}^{\text{in}}\right)\rightarrow\mathcal{L}\left(\mathcal{H}^{\text{out}}\right)$, where $\mathcal{L}\left(\mathcal{H}\right)$ denote all the linear operators on $\mathcal{H}$.

A separable state can always be written as a statistical mixture of product states:
\begin{equation}
\rho_{sep}=\sum_{i}p_{i}\rho_{A|i}\otimes\rho_{B|i}.
\end{equation}
All nonseparable states are entangled by definition.

A channel $\Phi_{\text{\textbf{EB}}}$ is entanglement-breaking (EB) if $\mathcal{I}^{A}\otimes\Phi_{\text{\textbf{EB}}}^{B}\left(\rho^{AB}\right)$ (equivalently, $\Phi_{\text{\textbf{EB}}}^{A}\otimes\mathcal{I}^{B}\left(\rho^{AB}\right)$) is separable for all initial states $\rho^{AB}\in\mathcal{S}\left(\mathcal{H}^{AB}\right)$. A map is EB if and only if it can be written in the following form
\citep{Ruskai2003,Horodecki2003}:
\begin{equation}
\Phi_{\text{\textbf{EB}}}\left(X\right)=\sum_{k}\tr\left[F_{k}X\right]\sigma_{k},
\end{equation}
where $F_{k}\geq0$ are positive semidefinite and $\sigma_{k}$ are fixed density states. If $\left\{ F_{k}\right\} $ form a positive operator valued measure (POVM) with $\sum_{k}F_{k}=\id$, then this channel is also trace-preserving. 

A channel $\Phi_{\text{\textbf{EA}}}$ is entanglement-annihilating (EA) if $\Phi_{\text{\textbf{EA}}}^{AB}\left(\rho^{AB}\right)$ is separable for all initial states $\rho^{AB}$ \citep{Moravcikova2010}. This can be extended to multipartite EA channels that destroy entanglement across $A|B|C|\cdots$ for any initial states $\rho^{ABC\cdots}$. There is no closed representation for entanglement-annihilating channels. For the interested reader, Ref. \citep{Filippov2012} gives  characterisations for two-qubit EA channels, and Ref. \citep{Filippov2013} gives necessary and sufficient conditions for bipartite EA channels.

Quantum discord is asymmetric. In this paper, system $A$ is the preferred system. Hence, quantum discord $D^{A}$ is defined as the difference between total correlations, given by the quantum mutual information $I\left(A:B\right)$, and classical correlations $J\left(B|A\right)$:
\begin{align}
D^{A}\left(\rho_{AB}\right) & =I\left(A:B\right)-J\left(B|A\right),\\
I\left(A:B\right) & =S\left(A\right)+S\left(B\right)-S\left(AB\right),\\
J\left(B|A\right) & =\max_{\Pi^{A}}\left[S\left(B\right)-\sum_{k}p_{a}S\left(\rho_{B|a}\right)\right].
\end{align}
where $S\left(X\right)=-\tr\left[\rho_{X}\log_{2}\rho_{X}\right]$ is the von Neumann entropy, $B$ has conditional states $\rho_{B|a}=\tr_{A}\left[\left(\Pi_{a}^{A}\otimes\id^{B}\right)\rho_{AB}\left(\Pi_{a}^{A}\otimes\id^{B}\right)\right]/p_{a}$ with probabilities $p_{a}=\tr\left[\left(\Pi_{a}^{A}\otimes\id^{B}\right)\rho_{AB}\left(\Pi_{a}^{A}\otimes\id^{B}\right)\right]$, and $\left\{ \Pi^{A}\right\} $ denotes a von Neumann measurement
\citep{Ollivier2001,Henderson2001}.

With $A$ as our preferred system zero-discord states are \emph{classical-quantum} (CQ), whose set we denote as $\mathbf{CQ}\left(\mathcal{H}^{AB}\right)$. CQ states can be written in the following form:
\begin{align}
\rho_{\text{\textbf{CQ}}} & =\sum_{k}p_{k}\ket{k}\bra{k}^{A}\otimes\rho_{B|k},
\end{align}
where $\left\{ \ket{k}\right\} $ is some orthonormal basis on $A$ and $\rho_{B|k}$ are density states. CQ states can also be written as
\begin{equation}
\rho_{\text{\textbf{CQ}}}=\sum_{ij}A_{ij}\otimes\ket{i}\bra{j}_{B},
\end{equation}
where $A_{ij}$ are mutually commuting normal operators and $\left\{ \ket{i}\right\} $ is any orthonormal basis on $B$ \citep{Guo2012}.

\section{Discord-breaking channels\label{sec:Discord-breaking-channels}}

Discord-breaking (DB) channels are applied locally in order to break discord between the local system and any external systems. However, unlike separable states, zero-discord states are asymmetric. There is a preferred subsystem that is classical while the remaining subsystems are nonclassical. Hence, \emph{different} channels $\Phi_{\text{\textbf{DB}}}$ are discord-breaking depending on whether they act on the first system $\Phi_{\text{\textbf{DB}}}^{A}\otimes\mathcal{I}^{B}\left(\rho^{AB}\right)$ (type A) or the second system $\mathcal{I}^{A}\otimes\Phi_{\text{\textbf{DB}}}^{B}\left(\rho^{AB}\right)$ (type B). We encapsulate this in the following theorem:
\begin{thm}
\label{thm:discord-breaking channels} Discord-breaking channels consist of two classes, type A and type B:

\begin{enumerate}[label=(\Alph*)]
\item A channel $\Phi_{\text{\textbf{DB}}}^{A}$ that acts on subsystem $A$ is discord-breaking type A if and only if it is a quantum-classical entanglement-breaking channel \citep{Guo2013,Yao2013}:
\begin{align}
\Phi_{\text{\textbf{DB(q-c)}}}^{A}\left(X\right) & =\sum_{k}\tr\left[F_{k}X\right]\ket{k}\bra{k}^{A},\label{eq:discord_breaking_type_A}
\end{align}
where $\left\{ F_{k}\right\} $ are positive semidefinite operators and $\left\{ \ket{k}\right\} $ is some orthonormal basis. If $\left\{ F_{k}\right\} $ is a POVM with $\sum_{k}F_{k}=\id$, then the channel is trace-preserving.\label{enu:DB_type_A}
\item A channel $\Phi_{\text{\textbf{DB}}}^{B}$ that acts on subsystem $B$ is discord-breaking type B if and only if it is a fixed point channel:
\begin{align}
\Phi_{\text{\textbf{DB(point)}}}^{B}\left(X\right) & =\tr\left[X\right]\sigma^{B},\label{eq:discord_breaking_type_B}
\end{align}
where $\sigma^B$ is a fixed density operator.\label{enu:DB_type_B}
\end{enumerate}

\end{thm}

For the proof of Theorem \ref{thm:discord-breaking channels}.\ref{enu:DB_type_A}, see Refs. \citep{Guo2013,Yao2013}, and especially Proposition 3 and 4 of Ref. \citep{Guo2013}.

The full proof of Theorem \ref{thm:discord-breaking channels}.\ref{enu:DB_type_B} is given in Appendix \ref{app:Breaking-discord-point}. Briefly, discord-breaking (type B) channels $\Phi_{\text{\textbf{DB}}}^{B}$ must be also entanglement breaking, which immediately allows us to consider a restricted set of channels. By considering various separable initial states on $AB$, we show that the output states have zero discord only when the conditional states on $B$ are identical, and thus $\Phi_{\text{\textbf{DB}}}^{B}$ must be a point channel $\Phi_{\text{\textbf{DB(point)}}}^{B}$. The image of $\mathcal{I}^{A}\otimes\Phi_{\text{\textbf{DB(point)}}}^{B}$ are always  product states, which have zero discord.

Discord-breaking type A channels are commutativity-creating channels: $\left[\Phi_{\text{\textbf{DB(q-c)}}}^{A}\left(\rho\right),\Phi_{\text{\textbf{DB(q-c)}}}^{A}\left(\sigma\right)\right]=0$ for all $\rho,\sigma$ \citep{Yao2013}. This is consistent with the notion that quantum discord arises due non-commutativity (on $A$) \citep{Guo2016}. General local channels on system $B$ cannot enforce commutativity on $A$, which leads to discord-breaking type B channels that simply destroy all correlations.

\begin{figure}
\begin{centering}
\includegraphics[width=0.75\columnwidth]{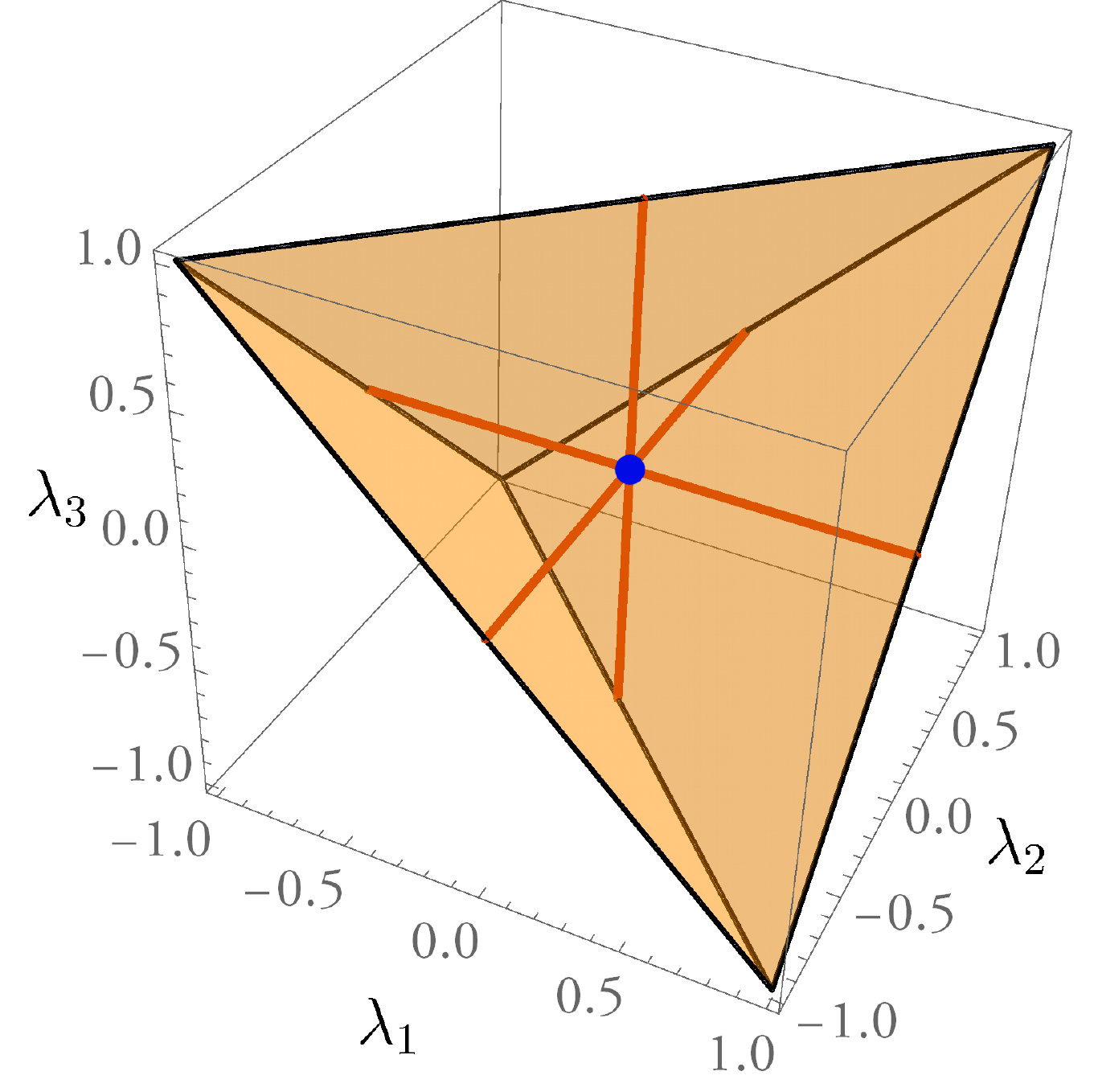}
\par\end{centering}
\caption{Unital qubit discord-breaking channels. The set of parameters $\left(\lambda_{1},\lambda_{2},\lambda_{3}\right)$ that correspond to a CPTP quantum channel for Eq. (\ref{eq:unital_qubit}) form a tetrahedron. If this channel acts on system $A$, then it is discord-breaking if and only if two  $\lambda_{i}=\lambda_{j}=0$ corresponding to the axes within the tetrahedron. If this channel acts on system $B$, then it is discord-breaking if and only if all $\lambda_{1}=\lambda_{2}=\lambda_{3}=0$ corresponding to the point at the center. Compare with Fig. 2 in Ref. \citep{Filippov2012}, where the set of unital qubit entanglement-breaking channels and -annihilating channels form nonzero volumes across the parameter space.\label{fig:Unital-qubit-discord-breaking}}
\end{figure}

Unital qubit channels can be reduced to the following representation (up to local unitaries that do not affect discord) \citep{King2001,Ruskai2002a}:
\begin{equation}
\hat{\mathcal{E}}=\begin{pmatrix}1 & 0 & 0 & 0\\
0 & \lambda_{1} & 0 & 0\\
0 & 0 & \lambda_{2} & 0\\
0 & 0 & 0 & \lambda_{3}
\end{pmatrix}.\label{eq:unital_qubit}
\end{equation}
in Fig. \ref{fig:Unital-qubit-discord-breaking}, we illustrate the local qubit discord-breaking channels. Type A channels lie on the axes of $\lambda_{1},\lambda_{2},\lambda_{3}$, and type B channels is the point at the origin $\lambda_{1}=\lambda_{2}=\lambda_{3}=0$. In contrast, entanglement-breaking and -annihilating unital qubit channels take a nonzero volume in the $\left(\lambda_{1},\lambda_{2},\lambda_{3}\right)$ paramater space (see Fig. 2 of Ref. \citep{Filippov2012}).

Before we conclude this section, we add some final notes on discord-breaking channels:
\begin{lem}
If $\Phi_{\text{\textbf{DB}}}^{A}$ is a discord-breaking type A channel, and $\mathcal{F}^{A}$ is any quantum channel, then $\Phi_{\text{\textbf{DB}}}^{A}\circ\mathcal{F}^{A}$ is also discord-breaking type A. If $\Phi_{\text{\textbf{DB}}}^{B}$ is a discord-breaking type B channel then $\Phi_{\text{\textbf{DB}}}^{B}\circ\mathcal{F}^{B}$ and $\mathcal{F}^{B}\circ\Phi_{\text{\textbf{DB}}}^{B}$ are also discord-breaking type B. \label{lem:DB_CPTP}
\end{lem}
\emph{Proof. }By the definition of discord-breaking channels, the composition $\Phi_{\text{\textbf{DB}}}^{X}\circ\mathcal{F}^{X}\left(\cdot\right)=\Phi_{\text{\textbf{DB}}}^{X}\left[\mathcal{F}^{X}\left(\cdot\right)\right]$, $X=A,B$ is discord-breaking. If $\Phi_{\text{\textbf{DB}}}^{B}\left(\cdot\right)=\tr\left[\cdot\right]\sigma^{B}$ has the fixed point $\sigma^{B}$, then $\mathcal{F}^{B}\circ\Phi_{\text{\textbf{DB}}}^{B}\left(\cdot\right)=\tr\left[\cdot\right]\mathcal{F}^{B}\left(\sigma^{B}\right)$ has fixed point $\mathcal{F}^{B}\left(\sigma^{B}\right)$. \hfill{}$\blacksquare$

In contrast, composition with $\mathcal{F}^{A}$ \emph{after} a type A channel, $\mathcal{F}^{A}\circ\Phi_{\text{DB}}^{A}$, is not discord-breaking in general. In order for $\mathcal{F}^{A}\circ\Phi_{\text{DB}}^{A}$ to be discord-breaking, $\mathcal{F}^{A}$ must be a channel that cannot create discord from zero-discord states, \emph{i.e.} it must be a discord-preserving channel. These are either completely decohering channels or isotropic channels \citep{Guo2013,Hu2012} (for qubits, there is an extra class of channels that are discord-preserving---see Theorem 2 of Ref. \citep{Guo2013}).

Lastly, entanglement-breaking channels form a convex set \citep{Horodecki2003}, but discord-breaking channels do not:
\begin{lem}
\label{lem:nonconvex_DB}The set of discord-breaking channels is not convex.
\end{lem}
\emph{Proof.} This is due to the nonconvexity of zero-discord states. For example, choose two discord-breaking type A channels with non-commuting output $\left\{ \ket{k}\bra{k}\right\} $ states. The convex sum will give separable discordant states in general (cf. Lemma \ref{lem:convex_CQ_sets}).  \hfill$\blacksquare$

This completes our study of discord-breaking channels---local channels that destroy quantum discord. In the following section, we consider their natural extension to nonlocal channels.

\section{Discord-annihilating channels\label{sec:Discord-annihilating-channels}}

Discord-annihilating (DA) channels are applied nonlocally on $AB$ to break the quantum discord within the system. We will find that the set of DA channels contain asymmetry in their definitions due to the asymmetry in classical-quantum states. Before we go on to characterise the exact form of discord-annihilating channels $\Phi_{\text{\textbf{DA}}}^{AB}$, we first require the following lemma, which describes the convex subsets of classical-quantum states: 
\begin{lem}
Convex subsets of classical-quantum states are $V=\text{conv}\left(W_{\text{\textbf{CQ}}}\right) \subset\mathbf{CQ}\left(\mathcal{H}^{AB}\right)$ where the states in $W_{\text{\textbf{CQ}}}$ share the following structure: \label{lem:convex_CQ_sets}
\begin{align}
W_{\text{\textbf{CQ}}} & \ni\sum_{i\in\mathcal{A}^{\text{BOTH}}}t_{i}\ket{\psi_{i}}\bra{\psi_{i}}\otimes R_{i}^{B}\nonumber \\
 & \phantom{\ni}+\sum_{i\in\mathcal{A}^{\text{FIXED}}}t_{i}\ket{\psi_{i}}\bra{\psi_{i}}\otimes\tilde{\sigma}_{i}^{B}\label{eq:convex_CQ_hybrid}\\
 & \phantom{\ni}+\sum_{i\in\mathcal{A}^{\text{POINT}}}t_{i}\tilde{\rho}_{A|i}\otimes R_{i}^{B},\nonumber 
\end{align}
where the index sets $\mathcal{A}^{\text{BOTH}}$, $\mathcal{A}^{\text{FIXED}}$, $\mathcal{A}^{\text{POINT}}$ are disjoint, the $t_{i}$ are probabilities $0\leq t_{i}\leq1$, $\sum_{i}t_{i}=1$, $\ket{\psi_{i}}\bra{\psi_{i}}\in\mathfrak{h}^{i}$ are orthonormal and span orthogonal subspaces, $\tilde{\rho}_{A|i}\in\mathfrak{h}^{i}\subset S\left(\mathcal{H}^{A}\right)$ such that all $\mathfrak{h}^{j}\cap\mathfrak{h}^{k}=\emptyset$ for $j\neq k$ are orthogonal subspaces, $\tilde{\sigma}_{i}^{B}\in S\left(\mathcal{H}^{B}\right)$ lives in a convex subset of states on $B$ (for each $i$), and $R_{i}^{B}\in S\left(\mathcal{H}^{B}\right)$ are fixed density states.
\end{lem}
The complete proof of Lemma \ref{lem:convex_CQ_sets} is given in Appendix \ref{app:Proof-of-convex_CQ}. It uses the following necessary condition for zero-discord states: $\left[\rho_{\text{\textbf{CQ}}}^{AB},\rho^{A}\otimes\id^{B}\right]=0$, where $\rho^{A}=\tr_{B}\left[\rho_{\text{\textbf{CQ}}}^{AB}\right]$ (Ref. \citep{Ferraro2010}, Prop. 1). By applying this condition onto the following state,
\begin{equation}
\rho^{AB}=p_{1}\sum_{i}q_{i}\ket{\psi_{i}}\bra{\psi_{i}}\otimes\sigma_{i}+p_{2}\sum_{j}r_{j}\ket{\phi_{j}}\bra{\phi_{j}}\otimes\varsigma_{j},\label{eq:convex_CQ_sum-1}
\end{equation}
we find that for each combination of $\left(i,j\right)$, either $\sigma_{i}-\varsigma_{j}=0$ or $\left[\ket{\psi_{i}}\bra{\psi_{i}},\ket{\phi_{j}}\bra{\phi_{j}}\right]=0$. This leads to components that are either mutually commuting on $A$ and/or have the same conditional state on $B$. This leads to the form in Eq. (\ref{eq:convex_CQ_hybrid}).

\begin{figure}
\begin{centering}
\includegraphics[width=1\columnwidth]{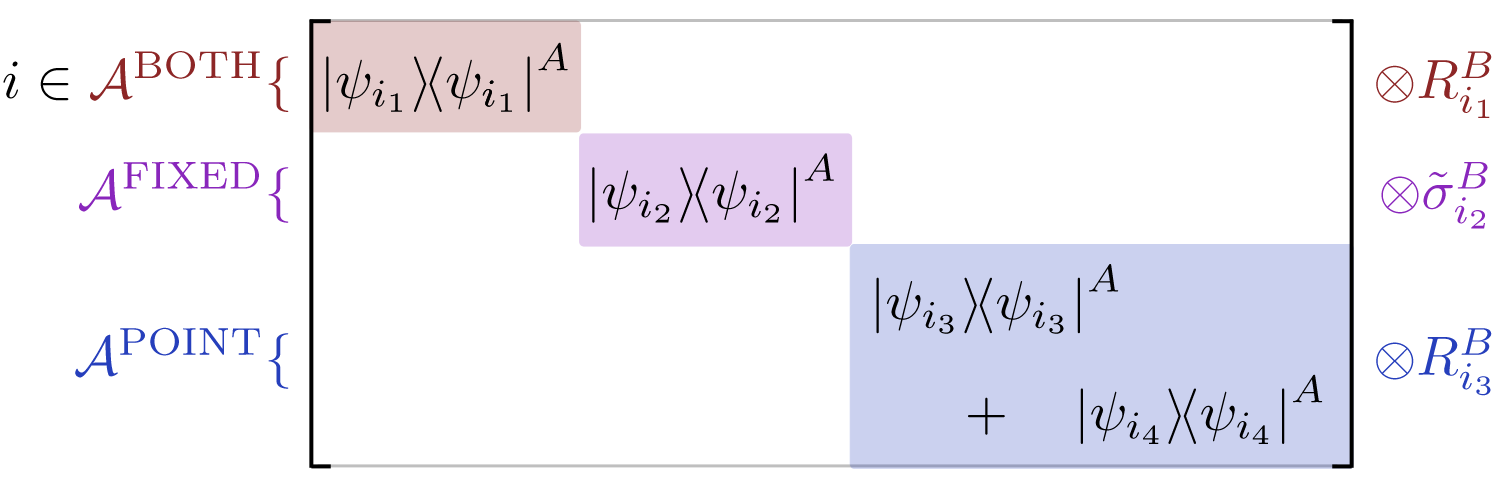}
\par\end{centering}
\caption{Depiction of states in a zero-discord convex subset, from Lemma \ref{lem:convex_CQ_sets} and Eq. (\ref{eq:convex_CQ_hybrid}). The state space on subsystem $A$ is broken into orthogonal subspaces indexed by $i\in\mathcal{A}^{\text{BOTH}}$, $\mathcal{A}^{\text{POINT}}$, $\mathcal{A}^{\text{FIXED}}$. Conditionally, the state on $A$ is either the orthonormal basis element $\left|\psi_{i}\right\rangle \left\langle \psi_{i}\right|^{A}$ for $i\in\mathcal{A}^{\text{BOTH}}\cup\mathcal{A}^{\text{POINT}}$, or is allowed to vary within an orthogonal subspace (with dimension two or greater) for $i\in\mathcal{A}^{\text{FIXED}}$. Conditionally, the state on $B$ is either the fixed point $R_{i}^{B}$ for $i\in\mathcal{A}^{\text{BOTH}}\cup\mathcal{A}^{\text{POINT}}$, or can vary within a convex subset of $B$ for $i\in\mathcal{A}^{\text{FIXED}}$. Whenever the basis on $A$ is fixed, the conditional state $\tilde{\sigma}^B_i$ on $B$ can vary; if the local basis on $A$ is not fixed, the conditional state on $B$ is fixed in order for an instantaneous orthonormal basis on $A$ to arise and fulfil the zero-discord condition.\label{fig:Convex_classical_quantum_subsets}}
\end{figure}

In Fig. \ref{fig:Convex_classical_quantum_subsets}, we depict the state structure from Eq. (\ref{eq:convex_CQ_hybrid}). Two classical-quantum states can be convexly combined into a new classical-quantum state only if they share the same state structure given in Eq. (\ref{eq:convex_CQ_hybrid}). The local conditional states on $A$ lie in orthogonal subspaces, and when these subspaces have dimension two or greater, the corresponding conditional state on $B$ \emph{must} be a fixed point. If all the orthogonal subspaces on $A$ correspond to the fixed orthonormal basis $\left\{ \ket{i}_{A}\right\} $ on $A$, then we have the following subset:
\begin{equation}
V_{\text{diag}A}^{\prime}=\left\{ \sum_{i}p_{i}\ket{i}\bra{i}_{A}\otimes\rho_{B|i}\left|\begin{array}{c}
\sum_{i}p_{i}=1,p_{i}\geq0,\\
\rho_{B|i}\in C\left\{ S\left(\mathcal{H}^{B}\right)\right\} 
\end{array}\right.\right\} ,
\end{equation}
where $C\left\{ S\left(\mathcal{H}^{B}\right)\right\} $ is a convex subset of $S\left(\mathcal{H}^{B}\right)$. If the orthogonal subspace on $A$ corresponds to the entire Hilbert space, then we have the following subset:
\begin{equation}
V_{\text{fixed}B}=\left\{ \rho_{A}\otimes R_{B}|\rho_{A}\in C\left\{ S\left(\mathcal{H}^{A}\right)\right\} \right\} ,\label{eq:convex_CQ_fixed_B}
\end{equation}
where $C\left\{ S\left(\mathcal{H}^{A}\right)\right\} $ is a convex subset of $S\left(\mathcal{H}^{A}\right)$, and $R_{B}\in S\left(\mathcal{H}^{B}\right)$ is a fixed density state.

We now present the main result for discord-annihilating channels:
\begin{thm}
A channel $\Phi_{\text{\textbf{DA}}}^{AB}$ is a discord-annihilating if and only if it can be written in the following form\label{thm:DA_channels}:
\begin{gather}
\Phi_{\text{\textbf{DA}}}^{AB}\left(\rho_{AB}\right)=\sum_{i}\Pi_{i}^{A}\otimes\Phi_{i}^{B}\left[\mathcal{E}^{AB}\left(\rho_{AB}\right)\right],\label{eq:DA_hybrid}
\end{gather}
where $\mathcal{E}^{AB}$ is a CPTP map, and the indices $i$ are divided into three disjoint sets $\mathcal{A}^{\text{BOTH}}$, $\mathcal{A}^{\text{POINT}}$, $\mathcal{A}^{\text{FIXED}}$. For a fixed orthonormal basis $\left\{ \ket{\psi_{i}}\right\}$, the projectors $\Pi_{i}^{A}$ either project into a one-dimensional subspace, or a multidimensional subspace, all of which are mutually orthogonal:
\begin{equation}
\Pi_{i}^{A}\left(\cdot\right)=\begin{cases}
\ket{\psi_{i}}\bra{\psi_{i}}^{A}\left(\cdot\right)\ket{\psi_{i}}\bra{\psi_{i}}^{A}, & i\in\mathcal{A}^{\text{BOTH}}\cup\mathcal{A}^{\text{FIXED}},\\
P_{i}^{A}\left(\cdot\right)P_{i}^{A}, & i\in\mathcal{A}^{\text{POINT}},
\end{cases}
\end{equation}
where $P_{i}$ are (higher-than-rank-one) projectors into orthogonal subspaces such that $\sum_{i}\ket{\psi_{i}}\bra{\psi_{i}}^{A}+\sum_{i}P_{i}=\id^{A}$. The conditional channels on $B$ are either point channels, or the identity channel:
\begin{equation}
\Phi_{i}^{B}\left(\cdot\right)=\begin{cases}
\Phi_{\text{\textbf{point}|}i}^{B}, & i\in\mathcal{A}^{\text{BOTH}}\cup\mathcal{A}^{\text{POINT}},\\
\mathcal{I}, & i\in\mathcal{A}^{\text{FIXED}}.
\end{cases}
\end{equation}
\end{thm}
The complete proof is given in Appendix \ref{app:Proof-of-DA_channels}. Briefly, suppose $\Phi_{\text{\textbf{DA}}}^{AB}$ is a discord-annihilating channel. It is a linear map, and since the set of all linear operators $\mathcal{L}\left(\mathcal{H}^{AB}\right)$ is convex, the image $\Phi_{\text{\textbf{DA}}}^{AB}\left(\mathcal{L}\left(\mathcal{H}^{AB}\right)\right)\subset\mathbf{CQ}\left(\mathcal{H}^{AB}\right)$ must also be convex. Thus, the image of $\Phi_{\text{\textbf{DA}}}^{AB}$ must be a convex subset of  zero-discord states which are precisely defined by Lemma \ref{lem:convex_CQ_sets}. Using the state structure defined in Lemma \ref{lem:convex_CQ_sets} we construct the most general channel structure that would lead to that fixed state structure and simplify til the form in Eq. (\ref{eq:DA_hybrid}) is achieved. Conversely, the channel in Eq. (\ref{eq:DA_hybrid}) directly leads to states given in Eq. (\ref{eq:convex_CQ_hybrid}) and hence is discord-annihilating.

\begin{figure}
\begin{centering}
\includegraphics[width=0.8\columnwidth]{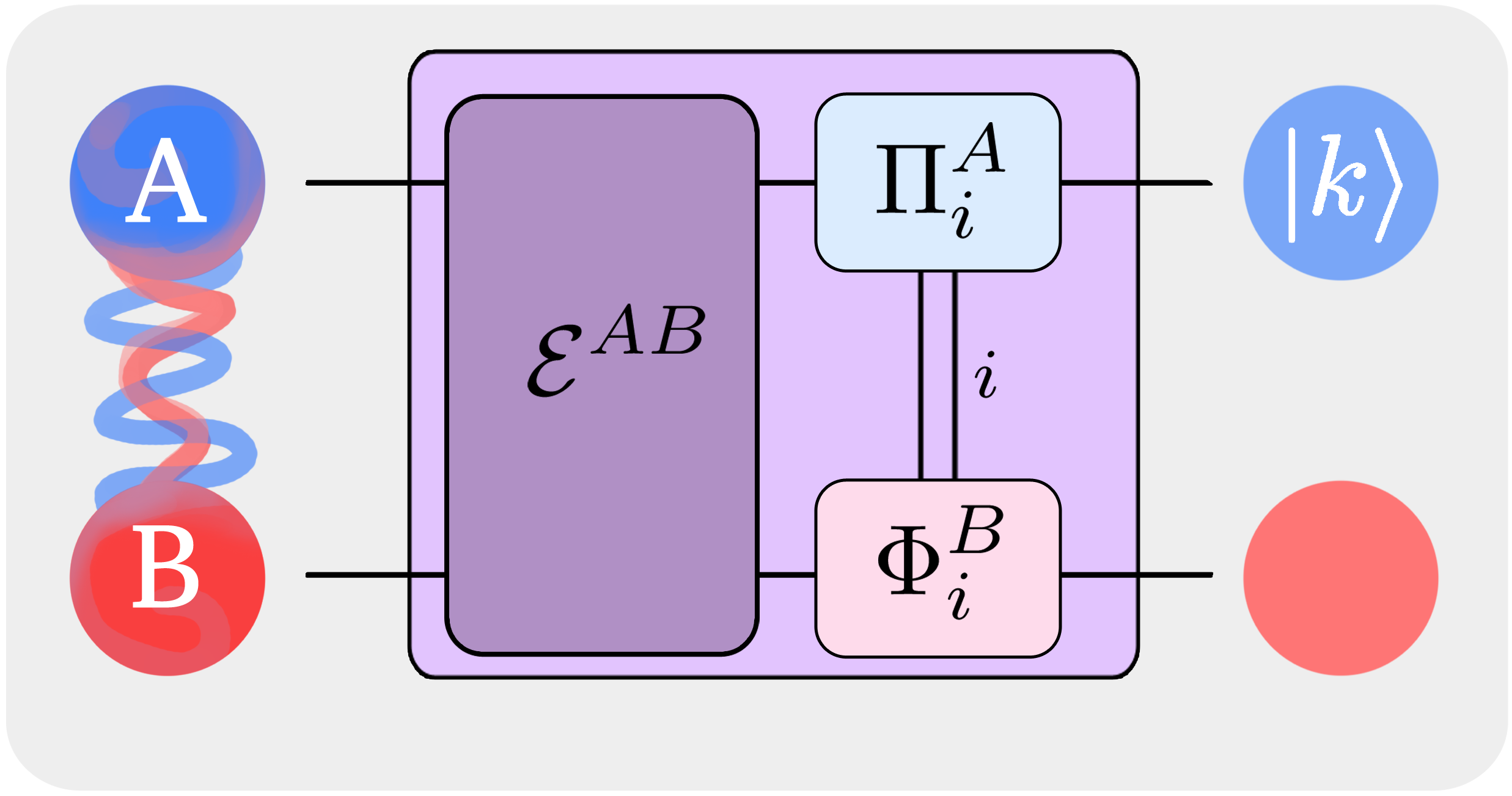}
\par\end{centering}
\caption{Discord-annihilating channels can be decomposed into initial arbitrary global dynamics $\mathcal{E}^{AB}$, followed by orthogonal projections  $\Pi_{i}^{A}$ on $A$ combined with conditional point channels or identity channels $\Phi_{i}^{B}$ on $B$. If the orthogonal projection is rank-two or higher, there must be a point channel on $B$. See Theorem \ref{thm:DA_channels} and Eq.
(\ref{eq:DA_hybrid}). \label{fig:Discord-annihilating-channels}}
\end{figure}

In Fig. \ref{fig:Discord-annihilating-channels}, we represent general discord-annihilating channels. They decompose into some initial arbitrary global dynamics $\mathcal{E}^{AB}$, followed by orthogonal-subspace-projections on $A$. When these subspaces have dimension greater than one, there is classical communication to $B$ leading to a conditional point channel on $B$. If the orthogonal subspaces are all one-dimensional, the channel reduces to a locally commutativity creating channel on $A$, with a fixed diagonal basis $\left\{ \ket{i}\bra{i}^{A}\right\}$, which corresponds to some arbitrary dynamics $\mathcal{E}^{AB}$ followed by a discord-breaking type A channel:
\begin{equation}
\Phi_{\text{\textbf{DA(fixed)}}}^{AB}\left(\rho_{AB}\right)=\left(\Phi_{\boldsymbol{\text{DB(q-c)}}}^{A}\otimes\mathcal{I}^{A}\right)\circ\mathcal{E}^{AB}.\label{eq:DA_fixed_diagonal_basis}
\end{equation}
If the orthogonal subspace corresponds to the entire Hilbert space of $A$, the channel decomposes into a point channel on $B$ and with arbitrary dynamics $\mathcal{E}^{AB}$:
\begin{align}
\Phi_{\text{\textbf{DA(point)}}}^{AB} & =\left(\mathcal{I}^{A}\otimes\Phi_{\boldsymbol{\text{DB(point)}}}^{B}\right)\circ\mathcal{E}^{AB}.\label{eq:DA_product}
\end{align}
If the discord-annihilating channel is local, the channel decomposition becomes even simpler:
\begin{cor}
Local discord-annihilating channels have form either $\Phi_{\text{\textbf{DB(q-c)}}}^{A}\otimes\mathcal{F}^{B}$ or $\mathcal{E}^{A}\otimes\Phi_{\text{\textbf{DB(point)}}}^{B}$ where $\Phi_{\text{\textbf{DB(q-c)}}}^{A}$ and $\Phi_{\text{\textbf{DB(point)}}}^{B}$ are discord-breaking type A and type B channels respectively, and $\mathcal{E}^{A},\mathcal{F}^{B}$ are any CPTP maps.\label{cor:product_DA_channels}
\end{cor}
\emph{Proof}. Following from Theorem \ref{thm:DA_channels}, if $\mathcal{E}^{A}\otimes\mathcal{F}^{B}$ is discord-annihilating, it must take the form (\ref{eq:DA_hybrid}), containing elements of commutativity creation on $A$ and fixed-point states on $B$. However, since $\mathcal{E}^{A}\otimes\mathcal{F}^{B}$ is product, the sum $\sum_{i}$ must only apply to $A$ or $B$---or that the channels must be independent of the sum, i.e. the commutativity creation on $A$ and the (single) fixed-point on $B$ must happen separately. This is only possible on all initial states if $\mathcal{E}^{A}$ is a discord-breaking type A channel; or (2) if $\mathcal{F}^{B}$ is a discord-breaking type B channel. \hfill$\blacksquare$

We would like to present an interesting fact about general discord-destroying channels:
\begin{prop}
If $\Phi^{AB}$ is discord-annihilating or discord-breaking, then $\det\hat{\Phi}^{AB}=0$, where $\hat{\Phi}^{AB}$ is any real representation of the channel.
\end{prop}
\emph{Proof. }If we have a linear map $T:\mathbb{R}^{n}\rightarrow\mathbb{R}^{n}$ and a set $A\subset\mathbb{R}^{n}$ that is Lebesgue measurable, then $\lambda\left(T\left(A\right)\right)=\left|\det T\right|\cdot\lambda\left(A\right)$, where $\lambda\left(X\right)$ denotes the Lebesgue measure of $X$ \citep{CharalambosD.PurdueUniversity1998}. Zero-discord states form a set of Lebesgue measure zero \citep{Ferraro2010}, while the set of all states has nonzero measure. We can represent quantum states in $\mathbb{R}^{n}$ by suitable mapping into an operator basis, and the quantum channels can be analogously transformed. Therefore, if we consider $T_{\text{\textbf{DA}}}$ discord-annihilating, and $A$ the set of all quantum states, the image of the map $T_{\text{\textbf{DA}}}\left(A\right)$ are a subset of zero-discord states, hence have measure zero: $\lambda\left(T_{\text{\textbf{DA}}}\left(A\right)\right)=0$. Hence, this implies that $0=\left|\det T_{\text{\textbf{DA}}}\right|\cdot\lambda\left(A\right)$. Since $\lambda\left(A\right)\neq0$, this implies that a necessary condition for discord-annihilating channels is $\det T_{\text{\textbf{DA}}}=0$.\hfill$\blacksquare$

Channels $\mathcal{E}$ with a zero determinant $\det\mathcal{E}$ are singular: hence the logarithm $\log\mathcal{E}$ does not \emph{strictly} exist and therefore $\mathcal{E}$ technically does not have a Lindblad form \citep{Wolf2008a}. Thus, discord-annihilating and discord-breaking channels are non-Markovian. Though, this is complicated by methods that can sometimes construct a logarithm on a suitable complex branch, and slight perturbations in the channel can be sufficient to produce similar channel that is Markovian  \citep{Wolf2008a,Cubitt2012}.  As zero discord states are nowhere-dense \citep{Ferraro2010}, this also suggests that discord-destroying channels are nowhere-dense in the space of quantum channels. Hence, slight perturbations of discord-destroying channels, i.e. almost-discord-annihilating channels, are Markovian.

Finally, we also have that:
\begin{lem}
The set of discord-annihilating channels is nonconvex.\label{lem:nonconvex_DA}
\end{lem}
This is because the set of zero-discord states is nonconvex, exactly as Lemma \ref{lem:nonconvex_DB}.

\section{Conclusion\label{sec:Conclusion}}

We examined various types of channels that destroy quantum discord: discord-breaking channels which act locally, and discord-annihilating channels which act nonlocally. Discord-destroying channels differ from  entanglement-breaking and -annihilating channels as quantum discord is inherently asymmetrical, and the set of zero-discord classical-quantum states is nonconvex. This led us to identity two classes of discord-breaking channels: \emph{type A} and \emph{type B}, corresponding to whether they act on subsystem $A$ or subsystem $B$ of the full bipartite system $AB$. Previous work had only identified discord-breaking type A channels, which correspond to quantum-classical entanglement breaking channels. We found that discord-breaking type B channels correspond to fixed-point channels. This form arises since local channels on $B$ cannot effect a preferred classical basis on $A$ for all states: point channels efficiently break all quantum and classical correlations.

We next examined discord-annihilating channels. We found that discord-annihilating channels contain a subtle interplay of measurement-projections on subsystem $A$ in conjunction with conditional fixed-point channels on $B$ whenever the projection has rank-two or greater. We noted that discord-annihilating and discord-breaking channels have zero determinant, which led to the subtle implication that these channels are  non-Markovian---though there are methods to describe them or their slight perturbations with Markovian dynamics.

Quantum discord is a vital correlation in numerous quantum applications. We have shown that discord-breaking and discord-annihilating channels take restricted forms, which we conjecture are nowhere-dense and zero-measure (analogous to the properties of the set of zero-discord states). Our work thus shows that quantum applications involving random channels are not at risk of complete loss of quantum correlations.

Conversely, the loss of quantum discord is an important component in the quantum-to-classical transition. For example, certain programs viewed the transition as involving the loss of quantum correlations with the spread of classical correlations \citep{Horodecki2015,Le2019,Zurek2009}. Our work opens up the potential to study the transition using discord-breaking and discord-annihilating channels and further restrictions thereof. We expect such a description to require a multipartite extension of discord-annihilating channels which we leave as a future exercise.

\section*{Acknowledgements}

This work was supported by the Engineering and Physical Sciences Research Council (Grant No. EP/L015242/1).

\appendix

\section{Proof of discord-breaking type B channels (Theorem \ref{thm:discord-breaking channels}.\ref{enu:DB_type_B}) \label{app:Breaking-discord-point}}

Discord-breaking channels must also be entanglement-breaking. Furthermore, they must work on any initial state. Therefore, consider some arbitrary initial state $\rho_{AB}=\sum_{i}p_{i}\rho_{A|i}\otimes\rho_{B|i}$. The action of the channel $\Phi_{\text{\textbf{DB}}}^{B}$ is:
\begin{align}
\mathcal{I}^{A}\otimes\Phi_{\text{\textbf{DB}}}^{B}\left(\rho_{AB}\right) & =\sum_{i,k}p_{i}\rho_{A|i}\otimes\tr\left[F_{k}^{B}\rho_{B|i}\right]R_{k}^{B}\\
 & \overset{!}{=}\sum_{j}q_{j}\ket{\psi_{j}}\bra{\psi_{j}}\otimes\rho_{B|j}^{\prime},\label{eq:app_type_B_must_have}
\end{align}
where the second line imposes the existence of a zero-discord decomposition, which will depend on the initial state. The various $\left\{ \ket{\psi_{j}}\right\} $ are orthonormal on $A$ and are equivalent to the spectral decomposition on $A$---since we have a local channel acting on a separable state, the state of $A$ remains locally unchanged. We can write the conditional states of $A$ in the $\left\{ \ket{\psi_{j}}\right\} $ basis:
\begin{align}
\rho_{A|i} & =\sum_{a,b}\braket{\psi_{a}|\rho_{A|i}|\psi_{b}}\ket{\psi_{a}}\bra{\psi_{b}},
\end{align}
then
\begin{multline}
\mathcal{I}^{A}\otimes\Phi_{\text{DB}}^{B}\left(\rho_{AB}\right)\\
=\sum_{a,b}\ket{\psi_{a}}\bra{\psi_{b}}\sum_{i}p_{i}\braket{\psi_{a}|\rho_{A|i}|\psi_{b}}\otimes\sum_{k}\tr\left[F_{k}^{B}\rho_{B|i}\right]R_{k}^{B}.
\end{multline}
To match with the form in Eq. (\ref{eq:app_type_B_must_have}), we require the following when $a\neq b$:
\begin{equation}
\sum_{i}p_{i}\braket{\psi_{a}|\rho_{A|i}|\psi_{b}}\left(\sum_{k}\tr\left[F_{k}^{B}\rho_{B|i}\right]R_{k}^{B}\right)\overset{!}{=}0.\label{eq:app_type_B_mid}
\end{equation}
Note that $\sum_{i}p_{i}\braket{\psi_{a}|\rho_{A|i}|\psi_{b}}=\braket{\psi_{a}|\rho_{A|i}|\psi_{b}}=0$ for $a\neq b$. The above equation (\ref{eq:app_type_B_mid}) holds if and only if $\sum_{k}\tr\left[F_{k}^{B}\rho_{B|i}\right]R_{k}^{B}$ has \emph{no }dependence on $i$, i.e. the output state on $B$ is a fixed point, which we will prove:

$\left(\Leftarrow\right)$ If the output state is a fixed point, then $\sum_{k}\tr\left[F_{k}^{B}\rho_{B|i}\right]R_{k}^{B}=\rho_{B}^{\prime}$ for some fixed $\rho_{B}^{\prime}$, and Eq. (\ref{eq:app_type_B_mid}) becomes
\begin{multline}
\sum_{i}p_{i}\braket{\psi_{a}|\rho_{A|i}|\psi_{b}}\left(\sum_{k}\tr\left[F_{k}^{B}\rho_{B|i}\right]R_{k}^{B}\right)\\
=\underbrace{\sum_{i}p_{i}\braket{\psi_{a}|\rho_{A|i}|\psi_{b}}}_{=0}\rho_{B}^{\prime}=0,
\end{multline}
as required.

$\left(\Rightarrow\right)$ Let $I_{SO}$ be the set of indices $i$ that give the same output, i.e. $\sum_{k}\tr\left[F_{k}^{B}\rho_{B|i}\right]R_{k}^{B}=\rho_{B,I_{SO}}^{\prime}$ for $i\in I_{SO}$, $\left|I_{SO}\right|\geq1$. Note that $\rho_{B|i}$ are fixed. Then Eq. (\ref{eq:app_type_B_mid}) becomes
\begin{align}
 & \sum_{i}p_{i}\braket{\psi_{a}|\rho_{A|i}|\psi_{b}}\left(\sum_{k}\tr\left[F_{k}^{B}\rho_{B|i}\right]R_{k}^{B}\right)\nonumber \\
 & =\left(\sum_{i\in I_{SO}}p_{i}\braket{\psi_{a}|\rho_{A|i}|\psi_{b}}\right)\rho_{B,I_{SO}}^{\prime}\\
 & \phantom{=}+\sum_{i\notin I_{SO}}p_{i}\braket{\psi_{a}|\rho_{A|i}|\psi_{b}}\rho_{B|i}^{\prime}\\
 & =\sum_{i\notin I_{SO}}p_{i}\braket{\psi_{a}|\rho_{A|i}|\psi_{b}}\left(-\rho_{B,I_{SO}}^{\prime}+\rho_{B|i}^{\prime}\right)\overset{!}{=}0.\label{eq:type_B_near_final}
\end{align}
where
\begin{equation}
    \sum_{i\in I_{SO}}p_{i}\braket{\psi_{a}|\rho_{A|i}|\psi_{b}} = -\sum_{i\notin I_{SO}}p_{i}\braket{\psi_{a}|\rho_{A|i}|\psi_{b}}.
\end{equation}
Eq. (\ref{eq:type_B_near_final}) must hold for all initial states, so we can choose a state where the term $\ket{\psi_{a}}\bra{\psi_{b}}$ only appears in two conditional $A$ states, $\rho_{A|i_{1}}$ and $\rho_{A|i_{2}}$ for example, such that $p_{i_{1}}\braket{\psi_{a}|\rho_{A|i_{1}}|\psi_{b}}+p_{i_{2}}\braket{\psi_{a}|\rho_{A|i_{2}}|\psi_{b}}=0$, $\braket{\psi_{a}|\rho_{A|i_{1},i_{2}}|\psi_{b}}\neq0$ and all others $i\neq i_{1},i_{2}$ have $\braket{\psi_{a}|\rho_{A|i}|\psi_{b}}=0$. 

If $i_{1},i_{2}\in I_{SO}$ then Eq. (\ref{eq:type_B_near_final}) is already satisfied. If only one $i_{1}\notin I_{SO}$ then the sum reduces to just one index and we must have $-\rho_{B,I_{SO}}^{\prime}=\rho_{B|i_{1}}^{\prime}$. If both $i_{1},i_{2}\notin I_{SO}$, and if we let $p_{i_{1}}=p_{i_{2}}=p$ due to freedom of initial state then Eq. (\ref{eq:type_B_near_final}) becomes:
\begin{gather}
\left\{ \begin{array}{c}
p\braket{\psi_{a}|\rho_{A|i_{1}}|\psi_{b}}\left(-\rho_{B,I_{SO}}^{\prime}+\rho_{B|i_{1}}^{\prime}\right)\\
+p\braket{\psi_{a}|\rho_{A|i_{2}}|\psi_{b}}\left(-\rho_{B,I_{SO}}^{\prime}+\rho_{B|i_{2}}^{\prime}\right)
\end{array}\right\} \overset{!}{=}0\\
\implies\rho_{B|i_{1}}^{\prime}\overset{!}{=}\rho_{B|i_{2}}^{\prime}.
\end{gather}
This procedure can be repeated for all pairs of $i\notin I_{SO}$ by suitable choice of initial states on $A$ and choices of $a,b$, leading to statement that output states with index $i\notin I_{SO}$ are the same: $\sum_{k}\tr\left[F_{k}^{B}\rho_{B|i}\right]R_{k}^{B}=\sigma_{B}^{\prime}$. Now, if we consider a more general initial state on $A$ again in Eq. (\ref{eq:app_type_B_mid}), 
\begin{align}
 & \sum_{i}p_{i}\braket{\psi_{a}|\rho_{A|i}|\psi_{b}}\left(\sum_{k}\tr\left[F_{k}^{B}\rho_{B|i}\right]R_{k}^{B}\right)\nonumber \\
 & =\sum_{i\in I_{SO}}p_{i}\braket{\psi_{a}|\rho_{A|i}|\psi_{b}}\rho_{B,I_{SO}}^{\prime}\nonumber \\
 & \phantom{=}+\sum_{i\notin I_{SO}}p_{i}\braket{\psi_{a}|\rho_{A|i}|\psi_{b}}\sigma_{B}^{\prime}\\
 & =\sum_{i\notin I_{SO}}p_{i}\braket{\psi_{a}|\rho_{A|i}|\psi_{b}}\left(-\rho_{B,I_{SO}}^{\prime}+\sigma_{B}^{\prime}\right)\overset{!}{=}0.\nonumber 
\end{align}
By choosing $\rho_{A|i}$ such that $\sum_{i\notin I_{SO}}p_{i}\braket{\psi_{a}|\rho_{A|i}|\psi_{b}}\neq0$, this is true if and only if $\rho_{B,I_{SO}}^{\prime}\overset{!}{=}\sigma_{B}^{\prime}$. Hence, all the output states on $B$ must be the same, and hence it is a point channel.

Lastly, point channels on entangled states will produce uncorrelated states, and hence zero-discord states.\hfill$\blacksquare$

\section{Proof of convex subsets of classical-quantum states (Lemma \ref{lem:convex_CQ_sets})\label{app:Proof-of-convex_CQ}}

Consider the convex combination of two composite CQ states:
\begin{equation}
\rho^{AB}=p_{1}\sum_{i}q_{i}\ket{\psi_{i}}\bra{\psi_{i}}\otimes\sigma_{i}+p_{2}\sum_{j}r_{j}\ket{\phi_{j}}\bra{\phi_{j}}\otimes\varsigma_{j}.\label{eq:convex_CQ_sum}
\end{equation}
Under the necessary condition $\left[\rho_{\text{\textbf{CQ}}}^{AB},\rho^{A}\otimes\id^{B}\right]=0$
\citep{Ferraro2010}, we require that 
\begin{equation}
\sum_{ij}q_{i}r_{j}\left[\ket{\psi_{i}}\bra{\psi_{i}},\ket{\phi_{j}}\bra{\phi_{j}}\right]\otimes\left(\sigma_{i}-\varsigma_{j}\right)\overset{!}{=}0.
\end{equation}
This holds if and only if, for each  $\left(i,j\right)$, either $\sigma_{i}-\varsigma_{j}=0$ or $\left[\ket{\psi_{i}}\bra{\psi_{i}},\ket{\phi_{j}}\bra{\phi_{j}}\right]=0$. To see this, note that the elements $\left\{ \left[\ket{\psi_{a}}\bra{\psi_{a}},\ket{\phi_{j}}\bra{\phi_{j}}\right]\right\} _{j}$ are linearly independent from the elements $\left\{ \left[\ket{\psi_{b}}\bra{\psi_{b}},\ket{\phi_{j}}\bra{\phi_{j}}\right]\right\} _{j}$ for $a\neq b$ \emph{unless} they are zero, in which case we have one of the aforementioned conditions. We can see this by choosing the basis $\ket{\psi_{a}}=\left(0,\ldots,1,\ldots,0\right)^{T}$
(with $1$ at position $a$), which leads to a commutator of form:
\begin{align}
\left[\ket{\psi_{a}}\bra{\psi_{a}},\ket{\phi_{j}}\bra{\phi_{j}}\right] & =\begin{bmatrix}0 & * & 0 & 0 \\
* & 0 & * & *\\
0 & * & 0 & 0\\
0 & * & 0 & 0
\end{bmatrix},
\end{align}
where the nonzero elements exist along either row $a$ or column $a$. Certain matrix elements only exist at a single given index $i\left(=a\right)$, hence there is linearly independence across the different indices $i$. Therefore, we can simplify our conditions to:
\begin{equation}
\sum_{j}r_{j}\left[\ket{\psi_{i}}\bra{\psi_{i}},\ket{\phi_{j}}\bra{\phi_{j}}\right]\otimes\left(\sigma_{i}-\varsigma_{j}\right)\overset{!}{=}0,\quad\forall i.
\end{equation}
The exact same argument can be applied to show that the commutators with different $j$, $\left[\ket{\psi_{i}}\bra{\psi_{i}},\ket{\phi_{j}}\bra{\phi_{j}}\right]$, are linearly independent  unless the commutator is zero, which regardless leads to the conditions that 
\begin{equation}
\left[\ket{\psi_{i}}\bra{\psi_{i}},\ket{\phi_{j}}\bra{\phi_{j}}\right]\otimes\left(\sigma_{i}-\varsigma_{j}\right)\overset{!}{=}0,\quad\forall i,j.
\end{equation}
This corresponds to either $\sigma_{i}-\varsigma_{j}=0$, $\ket{\psi_{i}}\bra{\psi_{i}}=\ket{\phi_{j}}\bra{\phi_{j}}$
or $\braket{\psi_{i}|\phi_{j}}=0$.

First, take the indices $i,j$ where $\ket{\psi_{i}}=\ket{\phi_{j}}$ \emph{and} $\sigma_{i}=\varsigma_{j}\equiv R_{i}^{B}$. Let $\mathcal{A}^{\text{BOTH}}$ be all the indices $i$ where this occurs, with the corresponding $j=j_{\text{BOTH}}\left(i\right)$. This leads to the first component of the sum $\rho^{AB}$ from Eq. (\ref{eq:convex_CQ_sum}):
\begin{equation}
\left[\rho^{AB}\right]_{\text{pt 1}}=\sum_{i\in\mathcal{A}^{\text{BOTH}}}\left(p_{1}q_{i}+p_{2}r_{j_{\text{BOTH}}\left(i\right)}\right)\ket{\psi_{i}}\bra{\psi_{i}}\otimes R_{i}^{B}.\label{eq:convex_CQ_part1}
\end{equation}
Since $\left\{ \ket{\psi_{i}}\right\} $ and $\left\{ \ket{\phi_{j}}\right\} $ respectively are orthonormal, the local state on $A$ of $\left[\rho^{AB}\right]_{\text{part 1}}$ is orthogonal to the local state on $A$ of the remainder $\rho^{AB}-\left[\rho^{AB}\right]_{\text{part 1}}$, \emph{i.e.}, they locally lie in orthogonal subspaces.

Next, consider the indices where $\ket{\psi_{i}}=\ket{\phi_{j}}$ and $\sigma_{i}\neq\varsigma_{j}$, where we define the index set $\mathcal{A}^{\text{FIXED}}$ for $i$, and $j=j_{\text{F}}\left(i\right)$. This leads to the second component:
\begin{equation}
\left[\rho^{AB}\right]_{\text{pt 2}}=\sum_{i\in\mathcal{A}^{\text{FIXED}}}\ket{\psi_{i}}\bra{\psi_{i}}\otimes\left(p_{1}q_{i}\sigma_{i}+p_{2}r_{j_{\text{F}}\left(i\right)}\varsigma_{j_{\text{F}}\left(i\right)}\right).
\end{equation}
Note that $\mathcal{A}^{\text{FIXED}}\cap\mathcal{A}^{\text{BOTH}}=\emptyset$
are disjoint, and similarly, the local components $\ket{\psi_{i}}\bra{\psi_{i}}$ in $\left[\rho^{AB}\right]_{\text{part 2}}$ are orthogonal to the local $A$ states in all the other parts.

Finally, we only have indices $i,j$ where  $\ket{\psi_{i}}\neq\ket{\phi_{j}}$. Some pairs may be orthogonal, but since they are not equal, they must be able to be written in terms of at least two other basis elements, e.g. $\ket{\phi_{j}}=\sum_{i}c_{i}\ket{\psi_{i}}$ where $c_{i}\neq0$ for at least two $i$'s. Then, for at least two $i$'s, we have $\left[\left[\ket{\psi_{i}}\bra{\psi_{i}},\ket{\phi_{j}}\bra{\phi_{j}}\right]\right]\neq0$ and hence we must have $\sigma_{i}=\varsigma_{j}=R_{i}^{B}$. Define the subsets of indices $i\in\mathcal{R}_{b}^{A}$ and $j\in\mathcal{R}_{b}^{B}$ where $\left[\left[\ket{\psi_{i}}\bra{\psi_{i}},\ket{\phi_{j}}\bra{\phi_{j}}\right]\right]\neq0$ that have the same state $\sigma_{i}=\varsigma_{j}=R_{b}^{B}$ on $B$. This defines the third component:
\begin{align}
\left[\rho^{AB}\right]_{\text{pt 3}}&=\sum_{b}\Bigl( \rho_{A|b} \Bigr)\otimes R_{b}^{B}.\\
\rho_{A|b} & \coloneqq \sum_{i\in\mathcal{R}_{b}^{A}}p_{1}q_{i}\ket{\psi_{i}}\bra{\psi_{i}}+\sum_{j\in\mathcal{R}_{b}^{B}}p_{2}r_{j}\ket{\phi_{j}}\bra{\phi_{j}}.
\end{align}
Note that the conditional $b$ local states on $A$ are orthogonal,
i.e. $\rho_{A|b} \in\mathfrak{h}^{b}$,
where $\mathfrak{h}^{b_{1}}\cap\mathfrak{h}^{b_{2}}=\emptyset$ for
$b_{1}\neq b_{2}$. This concludes all the indices, hence
\begin{equation}
\rho^{AB}=\left[\rho^{AB}\right]_{\text{pt 1}}+\left[\rho^{AB}\right]_{\text{pt 2}}+\left[\rho^{AB}\right]_{\text{pt 3}}.
\end{equation}
The conditional local states on $A$ exist on orthogonal subspaces. Any new states that can be added convexly must match this structure, for part 2 and part 3. In the case of part 1, if there is the same $R_{i}^{B}$ but not $\ket{\psi_{i}}\bra{\psi_{i}}$ or vice versa, then that component from part 1 will move to part 3 or part 2 respectively.

Hence, once the points in the set have been decided, the states in that set take form:
\begin{multline}
\sum_{i\in\mathcal{A}^{\text{BOTH}}}t_{i}\ket{\psi_{i}}\bra{\psi_{i}}\otimes R_{i}^{B}+\sum_{i\in\mathcal{A}^{\text{FIXED}}}t_{i}\ket{\psi_{i}}\bra{\psi_{i}}\otimes\tilde{\sigma}_{i}^{B}\\
+\sum_{i\in\mathcal{A}^{\text{POINT}}}t_{i}\tilde{\rho}_{A|i}\otimes R_{i}^{B},
\end{multline}
where the index sets $\mathcal{A}^{\text{BOTH}}$, $\mathcal{A}^{\text{FIXED}}$, $\mathcal{A}^{\text{POINT}}$ are disjoint, $t_i$ are the probabilities $0\leq t_{i}\leq1$, $\sum_{i}t_{i}=1$, $\tilde{\sigma}_{i}^{B}\in S\left(\mathcal{H}^{B}\right)$ live on convex subsets of states on $B$ for each $i$ and $\tilde{\rho}_{A|i}\in\mathfrak{h}^{i}\subset S\left(\mathcal{H}^{A}\right)$ where $\mathfrak{h}^{i}$ are convex and orthogonal:  $\mathfrak{h}^{b_{1}}\cap\mathfrak{h}^{b_{2}}=\emptyset$ for $b_{1}\neq b_{2}$ and $\mathfrak{h}^{b_{1}}$ is also orthogonal to all the explicitly written $\ket{\psi_{i}}\bra{\psi_{i}}$.\hfill$\blacksquare$

\section{Proof of discord-annihilating channels (Theorem \ref{thm:DA_channels})\label{app:Proof-of-DA_channels}}

Let $\Phi_{\text{\textbf{DA}}}^{AB}$ be a discord-annihilating channel. $\Phi_{\text{\textbf{DA}}}^{AB}$ is a linear map, and the set of all linear operators $\mathcal{L}\left(\mathcal{H}^{AB}\right)$ is convex. Hence the image  $\Phi_{\text{\textbf{DA}}}^{AB}\left(\mathcal{L}\left(\mathcal{H}^{AB}\right)\right)\subset\mathbf{CQ}\left(\mathcal{H}^{AB}\right)$ must also be convex and zero-discord. Such sets are precisely defined by Lemma \ref{lem:convex_CQ_sets}. Hence, we can write the output states as:
\begin{align}
\Phi_{\text{\textbf{DA}}}^{AB}\left(\rho_{AB}\right) & =\sum_{i\in\mathcal{A}^{\text{BOTH}}}\tr\left[\tilde{\mathcal{E}}_{i}^{AB}\left(\rho_{AB}\right)\right]\ket{\psi_{i}}\bra{\psi_{i}}\otimes R_{i}^{B}\nonumber \\
 & \phantom{=}+\sum_{i\in\mathcal{A}^{\text{FIXED}}}\ket{\psi_{i}}\bra{\psi_{i}}\otimes\tr_{A}\left[\tilde{\mathcal{F}}_{i}^{AB}\left(\rho^{AB}\right)\right]\label{eq:DA_channels_proof_start}\\
 & \phantom{=}+\sum_{i\in\mathcal{A}^{\text{POINT}}}\tr_{B}\left[\tilde{\mathcal{G}}_{i}\left(\rho_{AB}\right)\right]\otimes R_{i}^{B},\nonumber 
\end{align}
where $\tilde{\mathcal{E}}_{i}^{AB}$, $\tilde{F}_{i}^{AB}$ and $\tilde{G}_{i}$ are trace non-preserving quantum maps. Now, we can re-write each component:
\begin{widetext}
\begin{align}
\sum_{i\in\mathcal{A}^{\text{BOTH}}} & \tr\left[\tilde{\mathcal{E}}_{i}^{AB}\left(\rho_{AB}\right)\right]\ket{\psi_{i}}\bra{\psi_{i}}\otimes R_{i}^{B} = \sum_{i\in\mathcal{A}^{\text{BOTH}}}\ket{\psi_{i}}\bra{\psi_{i}}\otimes R_{i}^{B}\tr_{B}\left[\sum_{n}\bra{n}^{A}\tilde{\mathcal{E}}_{i}^{AB}\left(\rho_{AB}\right)\ket{n}^{A}\right]\\
 & =\sum_{i\in\mathcal{A}^{\text{BOTH}}}\ket{\psi_{i}}\bra{\psi_{i}}\tr_{B}\Bigl[\underbrace{\sum_{l}\ket{\psi_{l}}^{A}\left(\sum_{n,m}\bra{n}^{A}\tilde{\mathcal{E}}_{l}^{AB}\left(\rho_{AB}\right)\ket{n}^{AB}\right)\bra{\psi_{l}}^{A}}_{\eqqcolon \mathcal{E}^{AB}\text{ has no dependence on }i}\Bigr]\ket{\psi_{i}}\bra{\psi_{i}}\otimes R_{i}^{B}\\
 & =\sum_{i\in\mathcal{A}^{\text{BOTH}}}\ket{\psi_{i}}\bra{\psi_{i}}\tr_{B}\left[\mathcal{E}^{AB}\left(\rho_{AB}\right)\right]\ket{\psi_{i}}\bra{\psi_{i}}\otimes R_{i}^{B}\\
 & = \sum_{i\in\mathcal{A}^{\text{BOTH}}}\Pi_{i}^{A}\otimes\Phi_{\text{point}|i}^{B}\left[\mathcal{E}^{AB}\left(\rho_{AB}\right)\right], \quad \Pi_{i}^{A}\left(\cdot\right)\coloneqq \ket{\psi_{i}}\bra{\psi_{i}}\left(\cdot\right)\ket{\psi_{i}}\bra{\psi_{i}}, \Phi_{\text{point}|i}^{B}\left(\cdot\right)\coloneqq R_{i}^{B}\tr_{B}\left[\cdot\right],
\end{align}
\end{widetext}
where $\mathcal{E}^{AB}$ can be any CPTP map---we can write it in this form with an apparently specific $\mathcal{E}^{AB}$, but the channels of this form with a general $\mathcal{E}^{AB}$ will construct the state structure required. Similarly,
\begin{align}
 & \sum_{i\in\mathcal{A}^{\text{FIXED}}}\ket{\psi_{i}}\bra{\psi_{i}}\otimes\tr_{A}\left[\tilde{\mathcal{F}}_{i}^{AB}\left(\rho^{AB}\right)\right]\\
 & =\sum_{i\in\mathcal{A}^{\text{FIXED}}}\ket{\psi_{i}}\bra{\psi_{i}}^{A}\otimes\id^{B}\left(\mathcal{F}^{AB}\left(\rho^{AB}\right)\right)\ket{\psi_{i}}\bra{\psi_{i}}^{A}\otimes\id^{B}\\
 & =\sum_{i\in\mathcal{A}^{\text{FIXED}}}\Pi_{i}^{A}\otimes\mathcal{I}^{B}\left[\mathcal{F}^{AB}\left(\rho_{AB}\right)\right],
\end{align}
and
\begin{align}
 & \sum_{i\in\mathcal{A}^{\text{POINT}}}\tr_{B}\left[\tilde{\mathcal{G}}_{i}\left(\rho_{AB}\right)\right]\otimes R_{i}^{B}\\
 & =\sum_{i\in\mathcal{A}^{\text{POINT}}}P_{i}^{A}\left(\tr_{B}\left[\tilde{\mathcal{G}}_{i}\left(\rho_{AB}\right)\right]\right)P_{i}^{A}\otimes R_{i}^{B}\\
 & =\sum_{i\in\mathcal{A}^{\text{POINT}}}\Pi_{i}^{A}\otimes\Phi_{\text{point}|i}^{B}\left[\mathcal{G}^{AB}\left(\rho_{AB}\right)\right],
\end{align}
where $P_{i}^{A}\left(\tr_{B}\left[\tilde{G}_{i}\left(\rho_{AB}\right)\right]\right)P_{i}^{A}=\tr_{B}\left[\tilde{G}_{i}\left(\rho_{AB}\right)\right]$
since this component lives in the subspace projected into by $P_{i}^{A}$.

Then, we could define the following CPTP channel that gives the original dynamics from Eq. (\ref{eq:DA_channels_proof_start}):
\begin{align}
\Lambda^{AB}\left(\cdot\right) & =\sum_{i\in\mathcal{A}^{\text{BOTH}}}\ket{\psi_{i}}\bra{\psi_{i}}\mathcal{E}^{AB}\left(\cdot\right)\ket{\psi_{i}}\bra{\psi_{i}}\nonumber \\
 &  \phantom{=}+\sum_{i\in\mathcal{A}^{\text{FIXED}}}\ket{\psi_{i}}\bra{\psi_{i}}\mathcal{F}^{AB}\left(\rho_{AB}\right)\ket{\psi_{i}}\bra{\psi_{i}}\\
 & \phantom{=}+\sum_{i\in\mathcal{A}^{\text{POINT}}}P_{i}^{A}\mathcal{G}^{AB}\left(\rho_{AB}\right)P_{i}^{A}.\nonumber 
\end{align}
Hence
\begin{align}
\Phi_{\text{\textbf{DA}}}^{AB}\left(\rho_{AB}\right) & =\sum_{i\in\mathcal{A}^{\text{BOTH}}}\Pi_{i}^{A}\otimes\Phi_{\text{point}|i}^{B}\left[\Lambda^{AB}\left(\rho_{AB}\right)\right]\nonumber \\
 &\phantom{=} +\sum_{i\in\mathcal{A}^{\text{FIXED}}}\Pi_{i}^{A}\otimes\mathcal{I}^{B}\left[\Lambda^{AB}\left(\rho_{AB}\right)\right]\\
 &\phantom{=} +\sum_{i\in\mathcal{A}^{\text{POINT}}}\Pi_{i}^{A}\otimes\Phi_{\text{point}|i}^{B}\left[\Lambda^{AB}\left(\rho_{AB}\right)\right].\nonumber 
\end{align}
This is precisely Eq. (\ref{eq:DA_hybrid}), and it also holds for arbitary CPTP maps $\Lambda^{AB}$.

The converse direction of the Theorem is immediate.\hfill$\blacksquare$

\bibliographystyle{apsrev4-1_mod}
\bibliography{biblio}

\end{document}